\documentclass[11pt]{article}
  
\usepackage{amsmath,amssymb}

\usepackage{epsfig}  
\usepackage{graphicx}               
\usepackage{url}

\newcommand{\nc}{\newcommand}  

\nc{\beq}{\begin{equation}}  
\nc{\eeq}{\end{equation}}  
\nc{\beqa}{\begin{eqnarray}}  
\nc{\eeqa}{\end{eqnarray}}  
\nc{\bea}{\begin{eqnarray}}  
\nc{\eea}{\end{eqnarray}}  
\nc{\ra}{\rightarrow}  
\nc{\lsim}{\begin{array}{c}\,\sim\vspace{-21pt}\\< \end{array}}  
\nc{\gsim}{\begin{array}{c}\sim\vspace{-21pt}\\> \end{array}}  
\nc{\Tr}{{\rm Tr}}
\nc{\slsh}{\slash\hspace*{-0.22cm}}

\def\be{\begin{equation}}
\def\ee{\end{equation}}
\def\bea{\begin{eqnarray}}
\def\eea{\end{eqnarray}}

\newcommand{\Eref}[1]{Eq.~(\ref{#1})}

\title{  
\vspace*{-2.3cm}  
\begin{flushright}  
\normalsize{  
FERMILAB-PUB-10-209-T 
  }  
\end{flushright}  
\vspace{1.5cm}  
\Large  
\textbf{CP Violating Contribution to $\Delta \Gamma$ in the    $B_s$ System  from Mixing with a Hidden Pseudoscalar \\
}\vspace*{1.0cm}   
}

\author{Yang Bai$^a$
and Ann E. Nelson$^{b}$ 
\vspace{5mm}
\\
${}^{a}$\normalsize\emph{Theoretical Physics Department, Fermilab, Batavia, Illinois 60510} \\ \vspace{2mm}
${}^{b}$\normalsize\emph{Department of Physics, Box 1560, University of Washington, Seattle, WA 98195}
}

\date{}

\begin{document}  
\setcounter{page}{0}  
\maketitle  

\vspace*{1cm}  
\begin{abstract} 
Recent evidence for a $CP$ violating asymmetry in the semileptonic decays of $B_s$ mesons cannot be accommodated within the Standard Model.  Such an asymmetry can be explained by new physics contributions to $\Delta B=2$ components of either the mass matrix or the decay  matrix.  We show that  mixing with a hidden pseudoscalar meson  with a mass  around 5 GeV  can result in a new $CP$ violating contribution to the mixing and can resolve  several anomalies in this system including the width difference, the average width and the charge asymmetry. We also discuss the effects of the hidden meson on other $b$ physics observables, and present  viable decay modes for   the hidden meson. We make predictions for new decay channels of $B$ hadrons, which can be tested at the Tevatron, the LHC and B-factories.
\end{abstract}  
\thispagestyle{empty}  
\newpage  
  
\setcounter{page}{1}

\baselineskip18pt   

\section{Introduction}
\label{sec:intro}

In recent years, we have seen many new experimental measurements in the neutral $B$ meson system both from the B-factories and the Tevatron. While most   observations agree well with the Standard Model (SM), there are a few disagreements at the 2 or 3$\sigma$ level. Recently, the D$\emptyset$ collaboration has announced  evidence for a charge asymmetry in the number of like-sign dimuon events \cite{Abazov:2010hv}, which can be interpreted as a $CP$ violating asymmetry in $B_s$ meson oscillation rates and semileptonic decays. Such an asymmetry could result from a phase difference $\phi_s^{\rm sl}$ between  $\Gamma^{12}_s$ and $m^{12}_s$, 
\beqa\phi_s^{\rm sl} \equiv {\rm arg}(-m^{12}_s/\Gamma^{12}_s)\,,\eeqa 
where  $\Gamma^{12}_s$ is the off diagonal term in the $B_s$ decay matrix resulting from interference between $B_s$ and $\overline{B}_s$ decays, and $m^{12}_s$ is the  $\Delta B=2$ mass mixing term. In the SM, the dominant contribution to $m^{12}_s$ is   proportional to the weak Cabibbo-Kobayashi-Maskawa  (CKM) quark mixing matrix elements $(V^{\phantom{*}}_{tb}V^*_{ts})^2$ while $\Gamma^{12}_s$ is approximately proportional to $(V^{\phantom{*}}_{cb}V^*_{cs}+V^{\phantom{*}}_{ub}V^*_{us})^2-[8m_c^2/(3m_b^2)](V^{\phantom{*}}_{cb} V^*_{cs} + V^{\phantom{*}}_{ub} V^*_{us} )V^{\phantom{*}}_{cb} V^*_{cs}$. CKM unitarity constrains $V^{\phantom{*}}_{ts}V^*_{tb}$ to  equal $-V^{\phantom{*}}_{cs} V^*_{cb} - V^{\phantom{*}}_{us} V^*_{ub} $,  predicting \cite{Bigi:1987in, Lenz:2006hd}
\beqa
\phi_s^{\rm sl}({\rm SM)} \approx \frac{8\,m_c^2}{3\,m_b^2}\,\times\,\beta_s^{J/\psi \phi}({\rm SM)}\equiv \frac{8\,m_c^2}{3\,m_b^2}\,\times\,{\rm arg}\left(- \frac{V^{\phantom{*}}_{ts}V^*_{tb}}{V^{\phantom{*}}_{cs} V^*_{cb} }\right) \,=\,0.0042\pm0.0014\,,
\eeqa
therefore the SM prediction for the asymmetry is unmeasurably small, making the charge asymmetry an interesting  place to look for new physics. 

Several other observables in the $B_s$ system are in marginal disagreement with the SM. From the particle data group (PDG)~\cite{PDGMesons}, the average  lifetime of the neutral $B_s$ mesons is $\tau_{B_s}=1.472^{+0.024}_{-0.026}$~ps and the lifetime of the neutral $B_d$ mesons is $\tau_{B_d}=1.525\pm 0.009$~ps. The ratio of those two lifetimes is $\tau_{B_s}/\tau_{B_d}=0.965\pm 0.017$, which exhibits  a $1.8\sigma$ deviation from the SM prediction of $1.00\pm0.01$~\cite{Tarantino:2003qw,Gabbiani:2004tp}. Using the measured width of the $B_d$ (we assume that new physics does not modify the mixings in the $B_d$ system throughout this paper), the SM model prediction of the average  width of $B_s$ is 
\beqa\overline{\Gamma}_s({\rm SM})=0.654\pm 0.008~\mbox{ps}^{-1} ,\eeqa while the measured average  width is $\overline{\Gamma}_s = 0.680\pm{0.012}~\mbox{ps}^{-1}$.

The time-dependent $CP$ asymmetry in $B_s \rightarrow J/\psi\,\Phi$ decay also determines various mixing parameters in the $B_s$ mesons. Using the combined results from D$\emptyset$ and CDF~\cite{D0CDF28, SLACB} with 0-2.8~fb$^{-1}$ luminosity, the two extracted quantities are $\Delta \Gamma_s=0.154^{+0.054}_{-0.070}~\mbox{ps}^{-1}$ and $\beta_s^{J/\psi \phi}=0.39^{+0.18}_{-0.14}$, where $2\beta_s^{J/\psi \phi}$ is the $CP$ violating phase difference between the   mixing amplitude  and the decay amplitude, and 
$\Delta \Gamma_s$,  the width difference, is predicted by the SM to be 
\beqa\Delta \Gamma_s({\rm SM})=0.098\pm 0.024~\mbox{ps}^{-1}\ .\eeqa  The latest results from CDF with 2.8-5.2~fb$^{-1}$ luminosity are $\Delta \Gamma_s=(0.121\pm 0.051)~\mbox{ps}^{-1}$ and $\beta_s^{J/\psi \phi}=0.01\pm 0.17$~\cite{CDF52}. Combining the results, we have $\Delta \Gamma_s=(0.134\pm 0.039)~\mbox{ps}^{-1}$ and $\beta_s^{J/\psi \phi}=0.21\pm{0.12}$, differing  by a modest   $1.7\sigma$ from the SM.

The recent like-sign dimuon charge asymmetry $A^b_{\rm sl}$ of semileptonic $b$-hadron decays from D$\emptyset$ is $A^b_{\rm sl} = (-9.57\pm2.51\pm1.46)\times 10^{-3}$~\cite{Abazov:2010hv} for 6.1~fb$^{-1}$ luminosity. Using the SM predicted value of $a^d_{\rm sl} =  (-4.8^{+1.0}_{-1.2})\times 10^{-4}$ (we assume that the new physics contribution in the $B_d$ system is small) and combining with the explicit measurement $a^s_{\rm sl} = -(1.7\pm 9.1)\times 10^{-3}$~\cite{Abazov:2009wg},  the charge asymmetry for ``wrong-charge" semileptonic $B_s$-meson decay is $a^s_{\rm sl} \,=\, -(12.5\pm 4.8)\times 10^{-3}$. The SM prediction is $a^s_{\rm sl}({\rm SM})=(2.1\pm 0.6)\times 10^{-5}$~\cite{Lenz:2006hd}, which is off from the measured quantity by $2.6\sigma$. 

The three quantities $\Delta \Gamma_s$, $\phi_s^{\rm sl}$ and $a^s_{\rm sl}$ are not independent. We have the following  relation among them~\cite{Grossman:2009mn}
\beqa
a^s_{\rm sl}\,=\, \frac{|\Gamma_s^{12}|}{|m_s^{12}|}\,\sin{\phi_s^{\rm sl}} \, =\,\frac{\Delta \Gamma_s}{\Delta m_s}\,\tan{\phi_s^{\rm sl}} \  .
\label{eq:aslbetas}
\eeqa
The mass difference is measured very precisely: $\Delta m_s =17.78\pm 0.12~\mbox{ps}^{-1}$. Using the central value of $\Delta m_s$ and combining the measured values of $a^s_{\rm sl}$ and $\Delta \Gamma_s$, we find a good fit  with
\beqa
\Delta \Gamma_s\,=\,0.134\pm 0.031~\mbox{ps}^{-1}\,,     \qquad  \qquad  \tan\phi_s^{\rm sl}\,=\,-1.66\pm 0.64\,,
\eeqa
which are off from the SM predictions by $0.9\sigma$ and $2.6\sigma$, respectively. As we are anticipating new physics contributions to $\Gamma^{12}_s$ which do not necessarily  contribute to $B_s \rightarrow J/\psi\,\Phi$ decays we distinguish the fit to $\beta_s^{J/\psi \phi}$ from the fit to $\phi_s^{\rm sl}$. 

In constrast, the measured mass difference  in the $B_s$ system and the ratio of the mass differences between the $B_s$ and $B_d$ systems can be used to extract CKM matrix elements which are in fair  agreement with those extracted from other observables, although there is room for an ${\cal O}(20 \%)$ contribution to either mass difference from new physics~\cite{Buras:2009if,Sordini:2009gu}.

New physics can in principle contribute to both $\Gamma^{12}_s$ and $m^{12}_s$.   It is not possible to get a good fit to the dimuon asymmetry  obtained by D$\emptyset$   in terms of a new contribution to $m_s^{12}$ alone    \cite{Dobrescu:2010rh, Ligeti:2010ia}. When the SM value of $\Gamma^{12}_s$ is used in \Eref{eq:aslbetas}, fitting the leptonic asymmetry requires an  unphysical value for the phase $\phi_s^{\rm sl}$,
\be \sin\phi_s^{\rm sl}=-2.3\pm 1.3\ 
.\ee  A better fit is obtainable via a new contribution to $\Gamma^{12}_s$, which also can better fit the  modest deviations of $\overline{\Gamma}_s$ and $\Delta \Gamma_s$ from the SM. A general feature of models with new contributions to $\Gamma^{12}_s$ is that, in contrast with models which only modify $m^{12}_s$, the relation $\phi_s^{\rm sl}=-2\beta_s^{J/\psi \phi}$ does not hold~\cite{Kagan:2009gb}.

 Effective higher dimension operators offer a   general approach to any short distance new physics. Most attempts to explain the charge asymmetry have considered new short distance contributions~\cite{Chen:2010wv, Buras:2010mh, Deshpande:2010hy, Batell:2010qw, Chen:2010aq, Parry:2010ce, Ko:2010mn, King:2010np, Delaunay:2010dw}. In terms of $B_s$ and $\overline{B}_s$ and  decays into SM light particles, one can incorporate new physics  by  writing down effective operators and finding their allowed new decay channels.  This approach reveals the difficulty of obtaining large new contributions to $\Gamma^{12}_s$, and of obtaining a good fit to all available data.  New dimension 6 contributions to $\Delta B=2$   operators  contribute only to $m_s^{12}$. It is theoretically straightforward to construct theories which will produce such contributions which are comparable to those of the SM, since the SM $\Delta B=2$ operators occur only at the one-loop level and furthermore are proportional to small off-diagonal $V_{CKM}$ elements. Obtaining a new short distance contribution to $\Gamma^{12}_s$ requires $\Delta B=1$ operators which can contribute to $b$ quark decay. The SM $\Delta B=1$ operators are produced at tree level and so for new short distance physics to be important, there must  be new tree level contributions to  $b$ quark  decays  which are  comparable to the contribution of the  SM weak interactions. This is difficult to reconcile with the many successes of the SM   in predicting lifetimes and branching fractions of $B$ hadrons. Furthermore  any new particles contributing to $b$ quark  decays at the tree level also can produce $\Delta B=2$ operators at the one loop level, and so should give a large contribution to $m^{12}_s$.  A recent  analysis of the constraints on nonstandard $\Delta B=1$ operators which could contribute to $\Gamma_s^{12}$ has been given in Ref.~\cite{Bauer:2010dg}.

 It is however conceivable that new physics could also occur at longer distance scales, and would have escaped notice so far provided it is sufficiently weakly coupled to the SM particles.
In this paper, we  explore  in a concrete model how new, weakly coupled  physics at the  several GeV scale can give a large contribution to  $\Gamma^{12}_s$ while giving a contribution to $m^{12}_s$ which is smaller than that of the SM. In particular,  we assume that there is a new spin-zero particle, called $\zeta$, which can mix together with $B_s$ and $\overline{B}_s$. Although the amount of mixing could be small (otherwise a large modification on $\Delta m_s$ is anticipated), provided   $\zeta$  has a much larger width than the   $B_s$, $\Delta \Gamma_s$ can   be increased to match the experimental value even with a very small mixing between $\zeta$ and $B_s$, $\overline{B}_s$. As the amount needed to increase $\Delta \Gamma_s$ is comparable to the amount needed for the discrepancy in $\overline{\Gamma}_s$,   such a  model may explain both anomalies. Furthermore, if the mixing parameters between $\zeta$ and $B_s$, $\overline{B}_s$ contain new $CP$ violation phases  of order  unity, a large $\phi_s^{\rm sl}$ may  be obtained to explain the charge asymmetry. 

This new light scalar particle $\zeta$, which has a  weak coupling to the SM particles, has many possible origins from a model building point of view. In this paper, we take a purely phenomenological approach. We  treat this particle as a generic pseudoscalar, which for instance could be a Pseudo Nambu-Goldstone Boson (PNGB). For example, it may behave as a familon~\cite{Feng:1997tn} from some spontaneously broken  approximate family symmetry.  Or it may be a meson in a hidden sector, perhaps composite. The  $\zeta$ particle may decay directly into SM  particles or  into other hidden states, which then decay  into  SM particles.

To have new contributions to $\Delta \Gamma_s$, one should have $\zeta$ mixed with $B_s$ and $\overline{B}_s$. Such mixing can also modify $\Delta m_s$. Simply from perturbation theory, one can estimate that the modifications on the widths are proportional to the square of the mixing angles, while the modifications on the masses have an additional   factor proportional to the mass difference of $\zeta$ and $B_s$ or $\overline{B}_s$. So, without doing detailed calculations, if the anomalies can be explained by mixing with another state, the  contribution to $\Delta m_s$ can be reduced provided the new particle mass is close to the average $B_s$ meson mass: $M_\zeta \sim \overline{m}_{B_s}$.

Our paper is organized as follows. We will first describe the interactions of this new scalar field in Section~\ref{sec:interactions}, then we will diagonalize this three-particle system in Section~\ref{sec:diagonalize}. In Section~\ref{sec:fit}, we perform a $\chi^2$ based analysis to determine the best-fit region of the model parameter space. After that, we discuss various viable decay channels and conclude in Section~\ref{sec:conclusion}.

\section{Interactions of this New Scalar Field}
\label{sec:interactions}

In this section, we  will focus on flavor changing interactions of the new spinless particle  with $b$ and $s$ quarks  and leave its interactions with other particles for Section~\ref{sec:conclusion}.

For our analysis, it is convenient to take  $\zeta$ to interact with SM fermions dominantly through derivative couplings, as would be the case for a PNGB.  General  flavor changing interactions may be written as
\beqa
{\cal L}&=&\frac{1}{2}\partial_\mu \zeta\, \partial^\mu \zeta \,-\,\frac{1}{2}\,M_\zeta^2\,\zeta^2\,+\,
\frac{1}{F}\,\partial_\mu\,\zeta\,\overline{\psi}_i\,\gamma^\mu\,(g^{ij}_V\,+\,g_A^{ij}\,\gamma_5)\,\psi_j \,+\,h.c.\,+\,\cdots \,, \nonumber \\
&=&\frac{1}{2}\partial_\mu \zeta\, \partial^\mu \zeta \,-\,\frac{1}{2}\,M_\zeta^2\,\zeta^2\,-\,
\frac{i}{F}\,\zeta\,\overline{\psi}_i\,\left[g^{ij}_V(m_i - m_j)\,+\,g_A^{ij}(m_i+m_j)\,\gamma_5\right]\,\psi_j \,+\,h.c.\,+\,\cdots
\,.
\eeqa
Here, $M_\zeta$ is the PNGB mass; $\psi_j$ denotes  mass eigenstate SM fermions; and $F$ is a parameter which could be    the    spontaneous  symmetry breaking scale of some global symmetry.   The flavor-dependent couplings $g_V^{ij}$ and $g_A^{ij}$ are in general complex numbers. Other couplings could also exist, but will not be relevant for this part of our analysis. We will consider some other  interactions in Section~\ref{sec:fit} and \ref{sec:conclusion}. The general interaction  terms to  describe the off-diagonal couplings with second and third generation quarks are  
\beq
 -\frac{1}{F}\,\partial_\mu\,\zeta\,\overline{b}\,\gamma^\mu\,(g^{bs}_V\,+\,g_A^{bs}\,\gamma_5)\,s
  -\frac{1}{F}\,\partial_\mu\,\zeta\,\overline{t}\,\gamma^\mu\,(g^{tc}_V\,+\,g_A^{tc}\,\gamma_5)\,c \,+\,h.c. \,.
 \label{eq:operators}
\eeq
In  a model where $\zeta$ is related to the breaking of global flavor symmetries, we would anticipate $g^{bs}_{V,A} \sim g^{tc}_{V,A}$, if the up-type quarks and down-type quarks transform similarly. In principle, this new particle $\zeta$ can also couple to the first-generation quarks. We assume that such couplings are small and neglect them.

For the first operator in Eq.~(\ref{eq:operators}), we integrate it by parts and use the following matrix element
\beq
\partial_\mu\langle 0| \overline{b}\gamma^\mu\gamma_5 s (0) |B_s (p) \rangle = f_{B_s}\,m_{B_s}^2 \,,
\eeq
yielding mass mixing terms between $\zeta$ and $B_s$, $\overline{B}_s$ 
\beq
e^{i\alpha}\,f^2\,\zeta\,B_s  \,+\,e^{-i\alpha}\,f^2\,\zeta\,\overline{B}_s\,,
\eeq
with $\alpha \equiv {\rm arg}(g^{bs}_A)$ and $f^2\equiv |g^{bs}_A|\,f_{B_s}\,m^2_{B_s}/F$. With $f_{B_s} \approx 0.231\pm0.015$~GeV~\cite{Gamiz:2009ku}  and $m_{B_s}=5.3663\pm0.0006$~GeV, we have
\beqa
f\,=\,0.0026\,\times \left( \frac{F/|g^{bs}_A|}{10^6~\mbox{GeV}}    \right)^{-1/2}~\mbox{GeV}\,.
\label{eq:fandF}
\eeqa
We work in a basis where   $m^{12}_s$ is real in order to give physical, reparameterization invariant,  meaning to the phase $\alpha$. If $\alpha$ is not zero, a new source of $CP$-violation enters the $B_s$ and $\overline{B}_s$ system.  

The $\zeta$ field may decay into other light particles in its own hidden sector, or into SM particles. At this moment, we will simply assume it has a non-negligible width $\Gamma_\zeta$ and come back its   decays later. So, in the model we are considering, there are four parameters needed to compute the effects of mixing  with $\zeta$ on the $B_s$ system: $M_\zeta$, $\Gamma_\zeta$, $f$ and $\alpha$.

\section{Diagonalization of the Mass Matrix}
\label{sec:diagonalize}

The mass-squared matrix can be written in the basis  $(B_s, \overline{B}_s, \zeta)$ as
\beqa
\renewcommand{\arraystretch}{1.5}
M^2\,=\,\left(\begin{tabular}{c c c}
$m^2_{B_s}$  & $\Delta m_{B_s}m_{B_s}$ &  $e^{i\alpha} \,f^2$ \\
$\Delta m_{B_s}m_{B_s}$ & $m^2_{B_s}$ & $e^{-i\alpha}\,f^2$ \\
$e^{-i\alpha}\,f^2$ & $e^{i\alpha}\,f^2$ & $M^2_\zeta$
\\
\end{tabular}
\right) \,,
\eeqa
which can be diagonalized by the following unitary matrix as $UM^2 U^\dagger ={\rm diag}\{m_1^2, m_2^2, m_3^2\}$,
\beqa
\renewcommand{\arraystretch}{1.5}
U\,=\,\left(\begin{tabular}{c c c}
$\frac{e^{i\theta_{12}}}{\sqrt{2}}$  & $-\frac{e^{-i\theta_{12}}}{\sqrt{2}}$ &  $i\,\theta_{13}$ \\
$\frac{e^{i\theta_{12}}}{\sqrt{2}}$ & $\frac{e^{-i\theta_{12}}}{\sqrt{2}}$ & $\theta_{23}$ \\
$\frac{e^{i\theta_{12}}(i\,\theta_{13}-\theta_{23})}{\sqrt{2}}$ & $\frac{e^{-i\theta_{12}}(-i\,\theta_{13}-\theta_{23})}{\sqrt{2}}$ & $1$
\\
\end{tabular}
\right) \,.
\eeqa
Here, the three rotation angles are
\beqa
\theta_{12}&=&\frac{1}{2}\arctan{\left[ \frac{-\sin{(2\alpha)} }{\cos(2\alpha) - \frac{\Delta m_{B_s}\,m_{B_s}\,(M_\zeta^2-m_{B_s}^2 )}{f^4}}  \right]}     \,, \label{eq:theta12} \\
\theta_{13}&=& \frac{\sqrt{2}  f^2 \sin(\alpha+\theta_{12})}{m_{B_s}^2 - M_\zeta^2} \,,
\hspace{3cm}
\theta_{23}\,=\, \frac{\sqrt{2}  f^2 \cos(\alpha+\theta_{12})}{m_{B_s}^2 - M_\zeta^2} \,,
\eeqa
where we have assumed that $f^2 \ll |m^2_{B_s} - M_\zeta^2|$ and only kept the leading terms. From the above equations, one can see that when $f=0$, all three mixing angles are zero and the $\zeta$ is decoupled from the $B_s$ and $\overline{B}_s$ system. When $f^4 \sim \Delta m_{B_s}\,m_{B_s}\,|M_\zeta^2-m_{B_s}^2|$ and $\alpha = {\cal O}(1)$,     the mixing angle $\theta_{12}$ is of order unity. While for $f^4 \gg \Delta m_{B_s}\,m_{B_s}\,|M_\zeta^2-m_{B_s}^2|$, $\theta_{12}\approx -\alpha$ and $\theta_{13}\approx 0$.

In the diagonalized basis $(B_1, B_2, B_3)$, the three mass eigenvalues are
\beqa
m_{1}&=&
m_{B_s} - \frac{\Delta m_{B_s}}{2}\,\cos{(2\theta_{12})}\,-\,\frac{\theta_{13}^2\, (M_\zeta^2 - m^2_{B_S})}{2\,m_{B_s}}  \,,\nonumber \\
m_{2}&=&
m_{B_s} + \frac{\Delta m_{B_s}}{2}\,\cos{(2\theta_{12})}\,-\,\frac{\theta_{23}^2\, (M_\zeta^2 - m^2_{B_S})}{2\,m_{B_s}}  \,,\nonumber \\
m_{3}&=& M_\zeta + \frac{f^4}{M_\zeta \, (M_\zeta^2 - m^2_{B_S})}  \,.
\label{eq:massdifference}
\eeqa
Including the decay width matrix  and working in the basis $(B_1, B_2, B_3)$, we need to diagonalize the following effective Hamiltonian
\beqa
\renewcommand{\arraystretch}{1.5}
{\cal H} \,=\,\left(\begin{tabular}{c c c}
$m_1$  &  &   \\
 & $m_2$ &  \\
 &  & $m_3$\\
\end{tabular}
\right) 
\,-\,\frac{i}{2}
\,U\,\left(\begin{tabular}{c c c}
$\Gamma_s$  & $-\Gamma_{12}^s$  & 0  \\
$-\Gamma_{12}^{s *}$  & $\Gamma_s$ & 0 \\
0 & 0  & $\Gamma_\zeta$ \\
\end{tabular}
\right) \,U^\dagger \,.
\eeqa
Considering the relative phase between $M^s_{12}$ and $\Gamma^s_{12}$ is small in the SM, we neglect the phase of $\Gamma^s_{12}$ from now on. We also choose the phase convention such that $\Delta m_{B_s}$ and $\Gamma^s_{12}$ are both positive quantities. Noting that $\Gamma_s, \Gamma_{12}^s \ll \Delta m_{B_s}$, we use the ordinary perturbation theory to calculate the eigenvalues and eigenvectors. The three eigenvalues are calculated to be 
\beqa
\mu_{B_{s,L}}&=& m_1\,+\,{\cal O}\left(\frac{(\Gamma^s_{12})^2}{\Delta m_{B_s}}\right)\,-\,\frac{i}{2}\left[\Gamma_s+\Gamma_\zeta\, \theta_{13}^2 + \Gamma_{12}\,\cos(2\theta_{12})\right]
\,, \nonumber \\
\mu_{B_{s,H}}&=& m_2\,+\,{\cal O}\left(\frac{(\Gamma^s_{12})^2}{\Delta m_{B_s}}\right)\,-\,\frac{i}{2}\left[\Gamma_s+\Gamma_\zeta \, \theta_{23}^2 - \Gamma_{12}\,\cos(2\theta_{12})\right]
 \,,\nonumber \\
\mu_{\zeta^\prime}&=& m_3 \,-\,\frac{i}{2}\left[\Gamma_\zeta - \Gamma_\zeta (\theta^2_{13} + \theta^2_{23})\right] \,.
\eeqa
Neglecting terms suppressed by $1/(m_1 - m_3)$ and $1/(m_2 - m_3)$, the formulae of the mass eigenstates of the two lighter one in terms of flavor eigenstates are
\beqa
B_{s,L} &=& \frac{e^{i\,\theta_{12}}}{\sqrt{2}}\left(1 +\frac{\theta_{13}\theta_{23}\Gamma_\zeta - \sin(2\theta_{12}) \Gamma_{12} }{2(m_1 - m_2)} \right) \,B_s 
\,-\,\frac{e^{-i\,\theta_{12}}}{\sqrt{2}}\left(1 -\frac{\theta_{13}\theta_{23}\Gamma_\zeta - \sin(2\theta_{12}) \Gamma_{12} }{2(m_1 - m_2)} \right) \,\overline{B}_s  \nonumber  \\
&&\,+\,{\cal O}(\theta_{13},\theta_{23})\,\zeta \,, \nonumber \\
B_{s,H} &=& \frac{e^{i\,\theta_{12}}}{\sqrt{2}}\left(1 +\frac{\theta_{13}\theta_{23}\Gamma_\zeta - \sin(2\theta_{12}) \Gamma_{12} }{2(m_1 - m_2)} \right) \,B_s 
\,+\,\frac{e^{-i\,\theta_{12}}}{\sqrt{2}}\left(1 -\frac{\theta_{13}\theta_{23}\Gamma_\zeta - \sin(2\theta_{12}) \Gamma_{12} }{2(m_1 - m_2)} \right) \,\overline{B}_s  \nonumber \\
&&\,+\,{\cal O}(\theta_{13},\theta_{23})\,\zeta \,,
\eeqa
The mass and width differences of the heavy state and the light state are
\beqa
\Delta \overline{m}_{s}&=& m_H - m_L = \cos{(2\theta_{12})}\,\Delta m_{B_s} 
- \frac{(\theta_{23}^2- \theta_{13}^2)\, (M_\zeta^2 - m^2_{B_S})}{2\,m_{B_s}} 
 \,,\label{eq:massdifference2} \\
\Delta \overline{\Gamma}_s&=& \Gamma_L - \Gamma_H = \Gamma_\zeta (\theta_{13}^2 - \theta_{23}^2) \,+\,2\,\Gamma_{12}\,\cos(2\theta_{12})\,. \label{eq:widthdifference}
\eeqa
The average width is
\beq
\overline{\Gamma}_s\,=\,\Gamma_s\,+\,\frac{\theta^2_{13}+\theta^2_{23}}{2}\,\Gamma_\zeta\,.
\eeq
From Eqs.~(\ref{eq:massdifference2}) and (\ref{eq:widthdifference}), we can see that compared to the width difference the mass difference can be suppressed by an extra factor $(M_\zeta- m_{B_s})/m_{B_s}$ if $M_\zeta$ is close to $m_{B_s}$. This fact provides us the possibility of increasing the width difference without changing the mass difference too much.

Neglecting the effects of $\zeta$ in the $B_s$ and $\overline{B}_s$ oscillation, we use the traditional formula to calculate the charge asymmetry
\beqa
a^s_{\rm sl}&=&\frac{\Gamma(\overline{B}_s \rightarrow \mu^+ X)-\Gamma(B_s \rightarrow \mu^- X)}{\Gamma(\overline{B}_s \rightarrow \mu^+ X)+\Gamma(B_s \rightarrow \mu^- X)}  
= \frac{|\frac{p}{q}|^2 - |\frac{q}{p}|^2}{|\frac{p}{q}|^2 + |\frac{q}{p}|^2}  \nonumber \\
&=& \frac{2\,\theta_{13}\theta_{23}\Gamma_\zeta}{m_1-m_2} - \frac{2\sin(2\theta_{12}) \Gamma_{12} }{m_1 - m_2}  \nonumber \\
&=& -\frac{2\,\theta_{13}\theta_{23}\Gamma_\zeta}{\Delta\overline{m}_{s}} + \frac{2\sin(2\theta_{12}) \Gamma_{12} }{\Delta\overline{m}_{s}}
\,.
\label{eq:chargeasymmetry}
\eeqa
When $\alpha=0$ and no new $CP$ violating physics exists, we have $\theta_{12}=\theta_{13}=0$ and hence $a^s_{\rm sl}=0$. In order to compare to the observables both in the charge asymmetry and in the $B_s\rightarrow J/\psi\,\Phi$ decay, we calculate the phase $\phi_s^{\rm sl}$ and $\beta_s^{J/\psi\Phi}$ in our model
\beqa
\tan{\phi_s^{\rm sl}}&=&-\,\frac{2\,\theta_{13}\theta_{23}\Gamma_\zeta - 2\sin(2\theta_{12}) \Gamma_{12} }{\Gamma_\zeta (\theta_{13}^2 - \theta_{23}^2) \,+\,2\,\Gamma_{12}\,\cos(2\theta_{12})} \,, 
\\
\beta_s^{J/\psi\Phi}&=& -\theta_{12}\,.
\eeqa
%

\section{Fit to Observables}
\label{sec:fit}

In this section, we want to use this new model to fit the five observables in the neutral $B_s$ meson system. We first summarize the various experimental values and the SM predictions in Table~\ref{tab:data}.
\begin{table}[!ht]
\renewcommand{\arraystretch}{1.8}
\begin{center}
\begin{tabular}{c|cc}
\hline \hline
   & Experimental    & SM prediction      \\  \hline
$\Delta \overline{m}_s$   &  $(17.78\pm 0.12)~\mbox{ps}^{-1}$     &  $(19.6 \pm 2.2)~\mbox{ps}^{-1}$    \\ \hline
$\Delta \Gamma_s$   &  $0.134\pm 0.031~\mbox{ps}^{-1}$     &  $(0.098\pm 0.024)~\mbox{ps}^{-1}$    \\ \hline 
$\overline{\Gamma}_s$   &  $0.680\pm{0.012}~\mbox{ps}^{-1}$     &  $ (0.654\pm 0.008)~\mbox{ps}^{-1}$    \\ \hline 
$\tan{\phi_s^{\rm sl}}$   &  $-1.66\pm 0.64$     &  $0.0042 \pm 0.0014$    \\ \hline 
$\beta_s^{J/\psi\Phi}$   &  $0.21\pm 0.12$     &  $0.018 \pm 0.001$    \\ \hline 
   \hline
\end{tabular}
\end{center}
\caption{The experimental values and SM predictions for the five observables considered in this paper.}
\label{tab:data}
\end{table}%

We have four model parameters, $M_\zeta$, $f$, $\alpha$, $\Gamma_\zeta$, to fit the five observables, $\Delta \overline{m}_s$, $\Delta \Gamma_s$, $\overline{\Gamma}_s$, $\tan{\phi_s^{\rm sl}}$ and $\beta_s^{J/\psi\Phi}$. To quantify the goodness of fit from the new physics, we define the following $\chi^2$
\beqa
\chi^2\,=\,\sum^5_{i=1}\frac{(O_i^{\rm model}\,-\,O_i^{\rm exp})^2}{\sigma^2_{\rm SM} \,+\, \sigma^2_{\rm exp}} \,,
\eeqa
with $O_i$ represents the five observables. Neglecting new physics contributions or setting $f=0$, we have $\chi^2({\rm SM}) =14.0$, which indicates a large discrepancy between the SM and those five observables.

We first look at the case with an approximately massless PNGB. Since the two-body decay process $b\rightarrow s +\zeta$ from the spectator model is open, we should also consider constraints from the decay width of $B_d$. The experimental measured value is $\Gamma_{B_d} = 0.634\pm 0.004~\mbox{ps}^{-1}$. However, the SM prediction for this quantity has a large uncertainty. For example, the decay constant $f_{B_d}=190\pm13$~MeV~\cite{Gamiz:2009ku} from the lattice QCD calculation, which gives around $14\%$ uncertainty to $\Gamma_{B_d}$. To be conservative, we neglect other possible uncertainties and require the new physics contribution to $\Gamma_{B_d}$ to be less than $0.09~\mbox{ps}^{-1}$. Other B meson and B hadrons do not constrain the parameter space further, because the relative experimental errors of their widths are higher than $\Gamma_{B_d}$ and the theoretical errors are comparable. 

\begin{figure}[!ht]
\begin{center}
\includegraphics[width=0.48\textwidth]{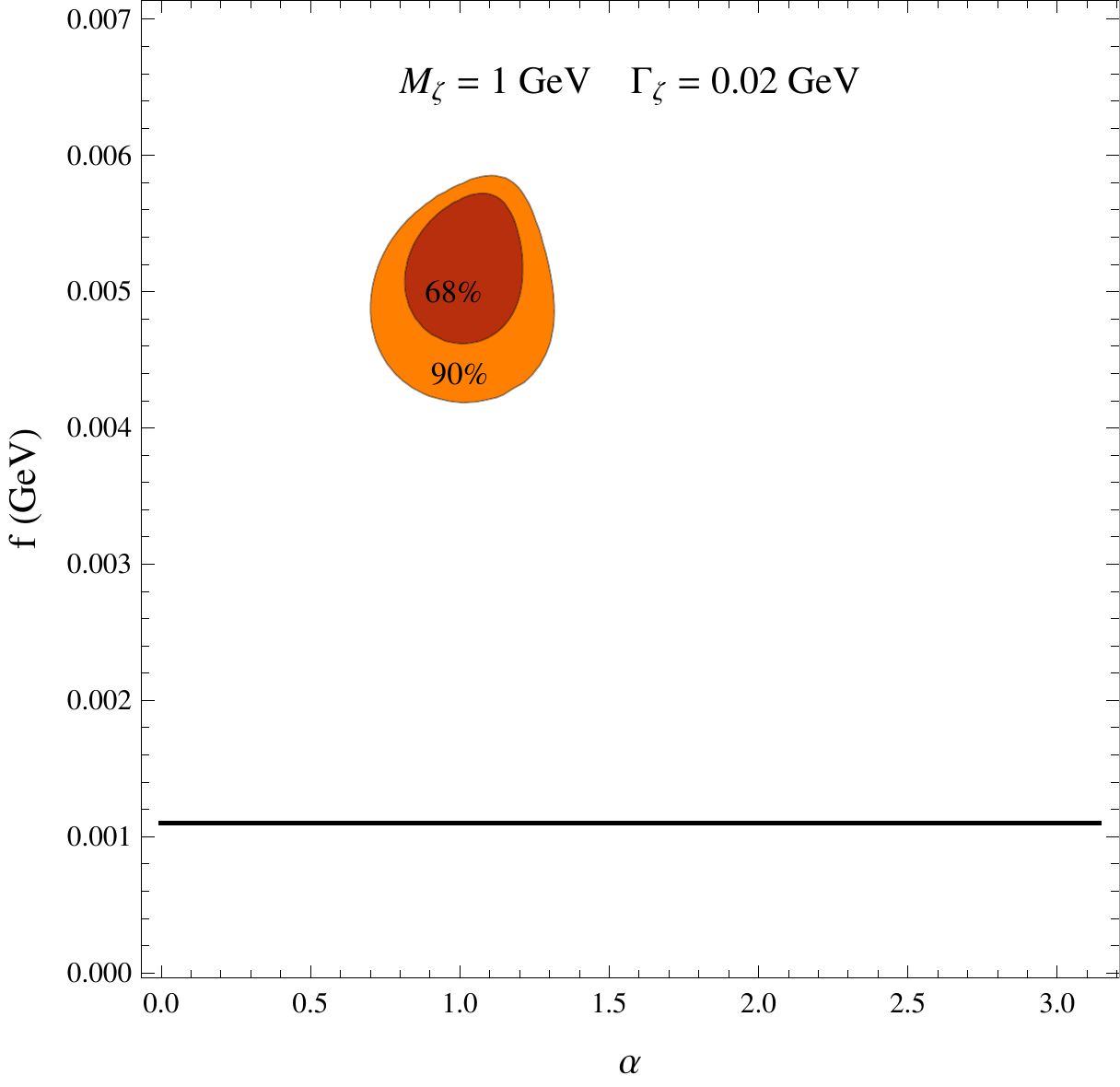} \hspace{2mm}
\includegraphics[width=0.48\textwidth]{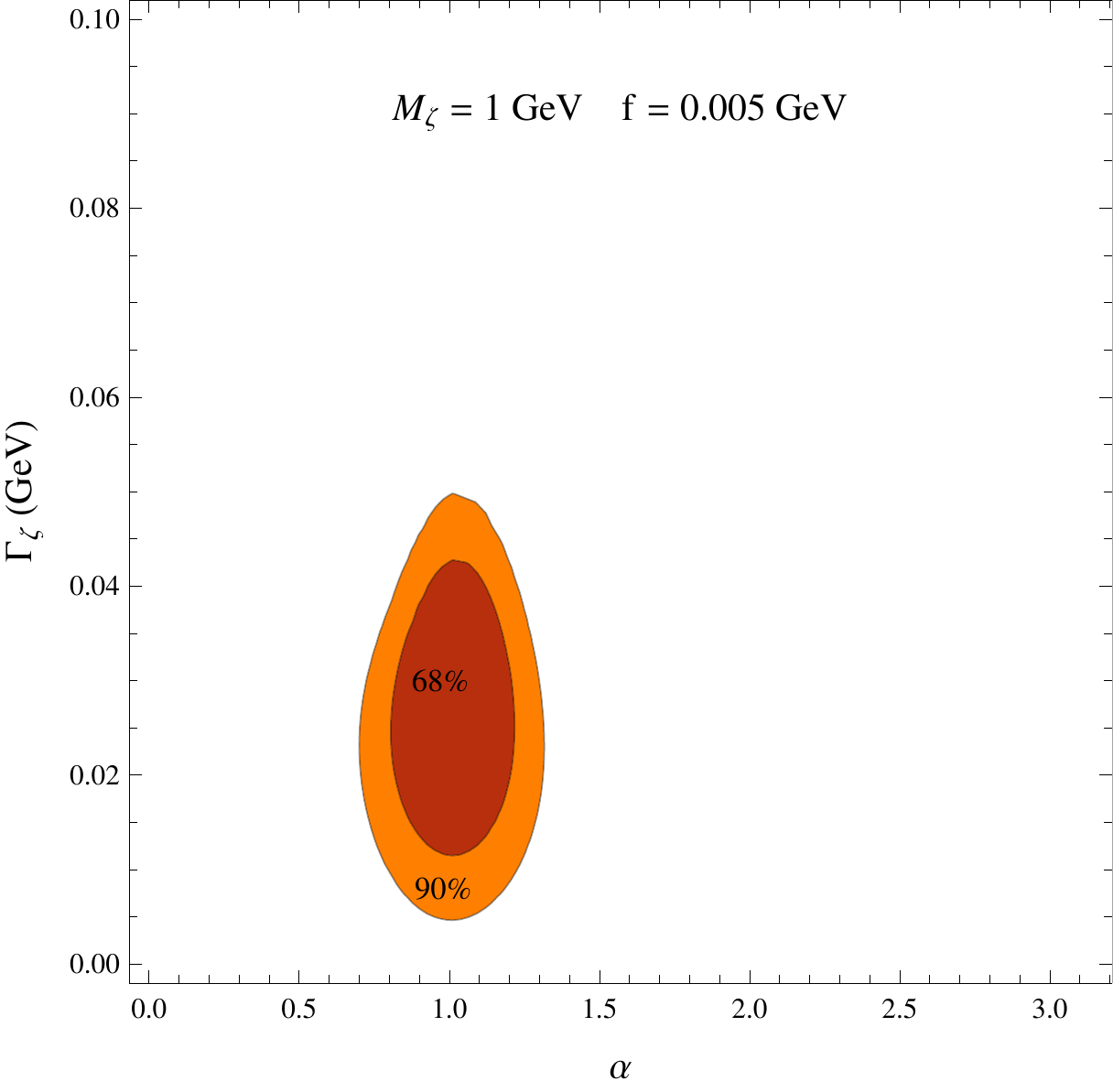}
\caption{Left panel: the best-fit region in the $f$ and $\alpha$ space for a light $\zeta$ mass and a fixed width. The global minimum has $\chi^2=2.0$. The two contours have $68\%$ and $90\%$ C.L., respectively. The region above the black solid line is ruled out due to decay width of $B_d$. Right panel: the same as the left panel but in the $\Gamma_\zeta$ and $\alpha$ plane.}
\label{fig:massless}
\end{center}
\end{figure}

For $M_\zeta < m_{B_d} - m_{K}$ with $m_{K}$ as the kaon mass, the two-body decay width of this channel is calculated to be
\beqa
\Gamma_{\rm spec}(b\rightarrow s+\zeta)\,\approx\,\frac{m_b^2\,|g_A^{bs}|^2}{16\,\pi\,F^2}\,\frac{(m^2_{B_d}-M_\zeta^2)^2}{m^3_{B_d}}\,=\,\frac{m_b^2\,f^4}{16\,\pi\,f^2_{B_s}\,m^4_{B_s}}\,\frac{(m^2_{B_d}-M_\zeta^2)^2}{m^3_{B_d}} \,.
\eeqa
Neglecting $M_\zeta$ and requiring $\Gamma_{\rm spec} < \delta \Gamma_{B_d} =0.09~\mbox{ps}^{-1}$, we have a constraint on $f < 1.1\times 10^{-3}$~GeV. Although a pretty good fit can be found for the approximately massless case, the allowed region is ruled out by the constraint from $\Gamma_{B_d}$. The best fit has $\chi^2=2.0$, and we present $68\%$ and $90\%$ contours around the best-fit region of our model parameters in Fig.~\ref{fig:massless}. We therefore conclude that the mass of the $\zeta$ field cannot be light compared with the $b$-quark mass. 

For $M_\zeta > m_{B_d} - m_{K}$, the two-body decay  channels are forbidden, but there are still three-body decay  channels open. The three-body decay  width is related to the width $\Gamma_\zeta$ because an off-shell $\zeta$ mediates the three-body decay. To be concrete, we assume that $\zeta$ can decay into two light scalar fields $a$ via the interaction $\kappa\,\zeta\,a^2/2$, where $\kappa$ has mass dimension one. The light scalar field $a$ can subsequently decay back into SM particles. Detailed discussions about possible decay  channels will be presented in Section~\ref{sec:conclusion}. We have the two-body decay width $\Gamma_\zeta$ as
\beqa
\Gamma_\zeta (\zeta\rightarrow 2\,a)\,=\,\frac{|\kappa|^2}{32\,\pi\,M_\zeta}\,.
\label{eq:phidecay}
\eeqa

\begin{figure}[!ht]
\begin{center}
\includegraphics[width=0.6\textwidth]{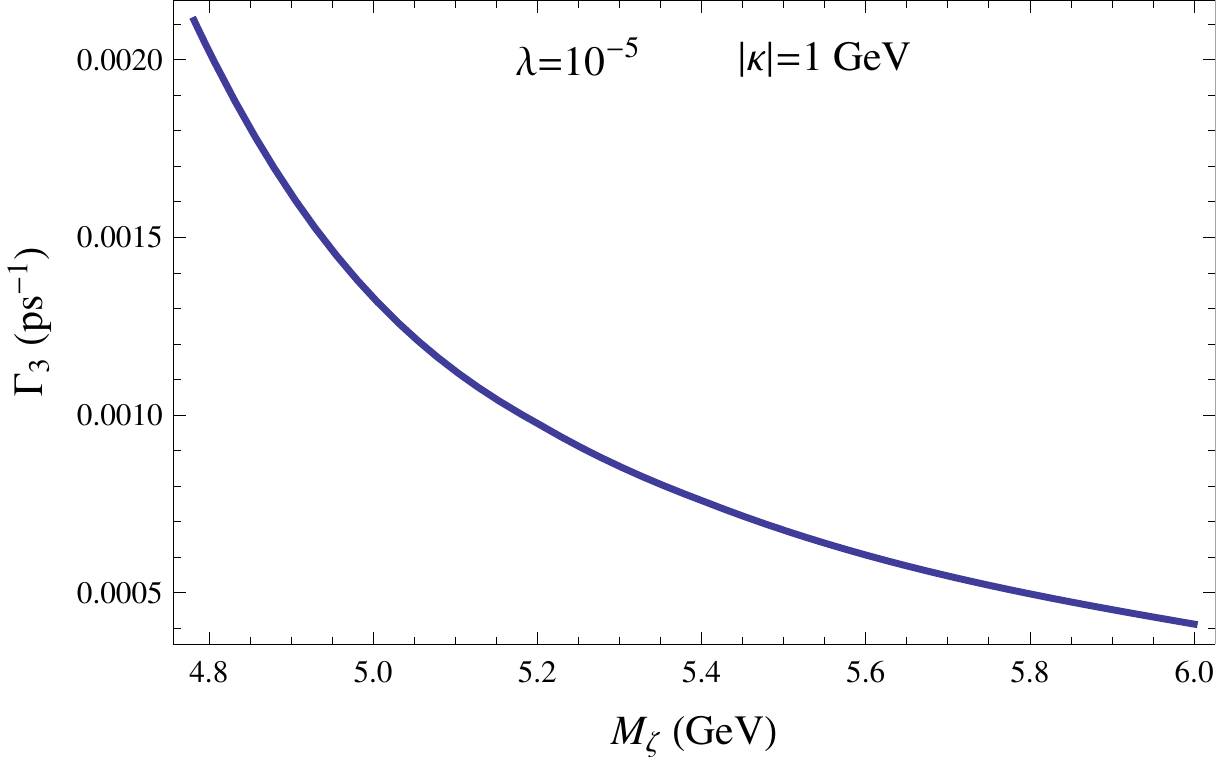} 
\caption{The three-body decay width of ${B_d}\rightarrow K + a + a$ for $\lambda=10^{-5}$ and $|\kappa|=1$~GeV.}
\label{fig:calcthreebody}
\end{center}
\end{figure}

Defining the Yukawa coupling $\lambda\equiv m_b\,|g^{bs}_A|/F$ and neglecting the mass of $a$, the three-body decay width, $\Gamma_3({B_d}\rightarrow K + a + a)$ through an off-shell $\zeta$, is calculated as
\beq
\Gamma_3 \,=\, \frac{\lambda^2\,|\kappa|^2}{32\,\pi^3}\,S\left(M_\zeta, m_{B_d}, m_K \right)\,.
\eeq
%
%
%
For $\lambda=10^{-5}$ and $|\kappa|=1$~GeV, we calculate this decay width using Calchep~\cite{Pukhov:2004ca} and show it in Fig.~\ref{fig:calcthreebody}.

Using Eq.~(\ref{eq:phidecay}) and the relations $f^2\equiv |g^{bs}_A|\,f_{B_s}\,m^2_{B_s}/F$ and $\lambda\equiv m_b\,|g^{bs}_A|/F$, we have the three-body decay width as a function of $f$ and $\Gamma_\zeta$
\beq
\Gamma_3 \,=\, \frac{f^4\,m_b^2\,M_\zeta\,\Gamma_\zeta}{\pi^2\,f_B^2\,m^4_{B_s}}\,S\left(M_\zeta, m_{B_d}, m_K\right)\,.
\label{eq:threebodywidth}
\eeq
In the following, we will impose the constraint $\Gamma_{3} < \delta \Gamma_{B_d} =0.09~\mbox{ps}^{-1}$.

\begin{figure}[!hbt]
\begin{center}
\includegraphics[width=0.48\textwidth]{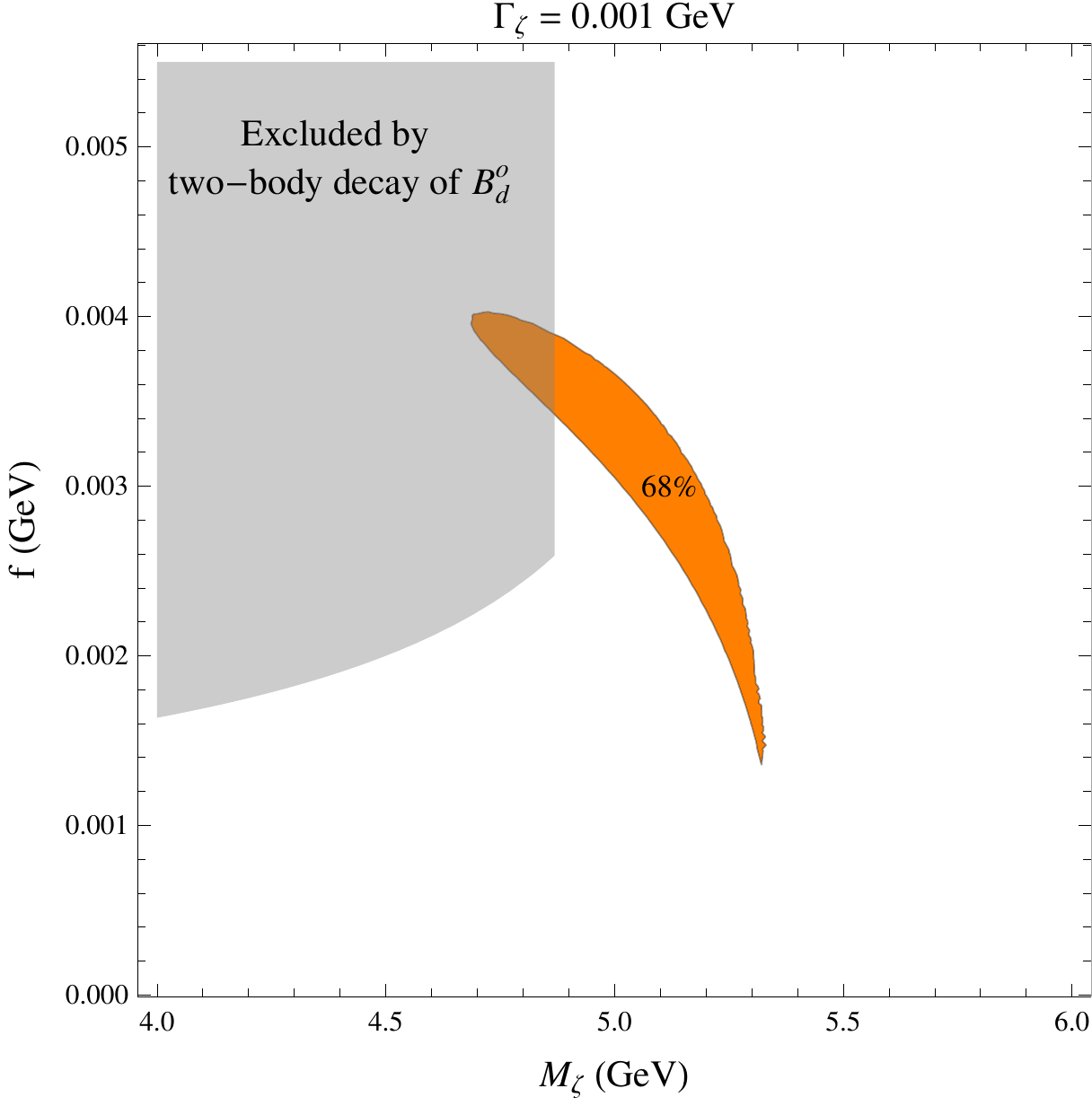} \hspace{2mm}
\includegraphics[width=0.48\textwidth]{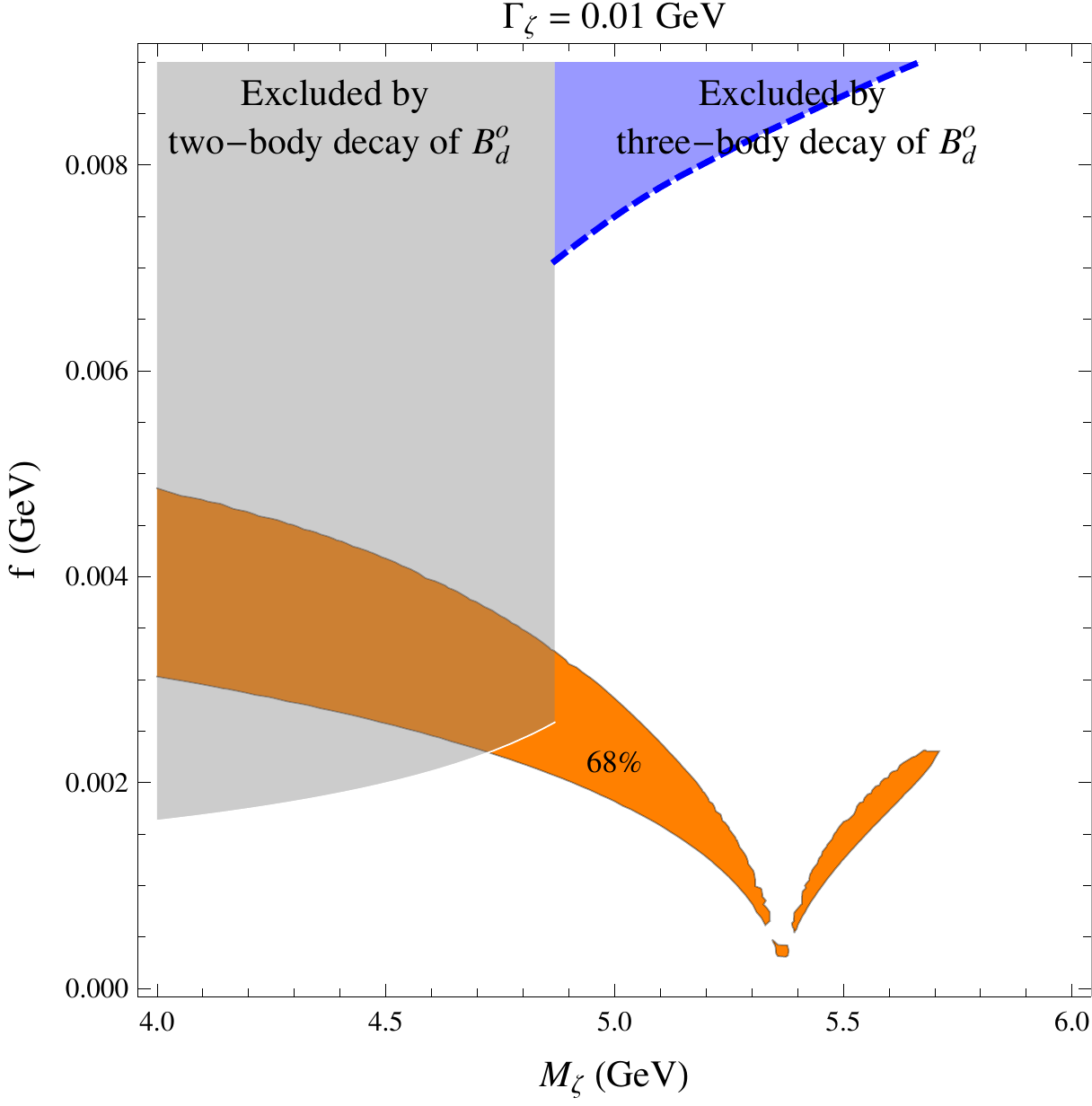}
\caption{Left panel: the best-fit region in the $M_\zeta$ and $f$ space for a fixed width $\Gamma_\zeta=0.001$~GeV. The orange contour has $68\%$ C.L. after minimizing $\chi^2$ in terms of $\alpha$. The best fit has $\chi^2=2.0$. The gray region is ruled out by the two-body decay width of $B_d$ when $M_\zeta < m_{B_d} - m_K$. Three-body decays do  not rule out the best-fit region. Right panel: the same as the left panel but for $\Gamma_\zeta=0.01$~GeV. The best fit has $\chi^2=5.4$. The blue region is excluded by requiring the three-body decay width to be below the error of $\delta\Gamma_{B_d} =0.09~\mbox{ps}^{-1}$.}
\label{fig:massiveWidth}
\end{center}
\end{figure}

We present the allowed parameter space in Fig.~\ref{fig:massiveWidth}. In the left panel, we first fix the decay width as $\Gamma_\zeta =0.001$~GeV, and then calculate the total $\chi^2$ by marginalizing $\alpha$. We have found a region of parameter space providing a much better fit to the five quantities in Table~\ref{tab:data} than from the SM. The best fit can have $\chi^2=2.0$. The right panel is similar to the left one but with $\Gamma_\zeta =0.01$~GeV. As can be seen from those plots, the  $\zeta$ is preferred to have a mass close to the $B_s$ meson. This is because when $M_\zeta \sim m_{B_s}$, one can have   fairly large changes to the quantities $\Delta \overline{\Gamma}_s$ and $a^s_{\rm sl}$ without a large contribution to $\Delta \overline{m}_s$. Comparing those two plots, one can see that the plot with a larger $\Gamma_{\zeta}$ has more parameter space ruled out by the three-body decay width, which can be understood simply from Eq.~(\ref{eq:threebodywidth}). To illustrate the goodness of our fit, we report the results for one point of our parameter space
\beqa
M_\zeta = 5.2~\mbox{GeV}\,, \qquad f = 0.0023~\mbox{GeV}\,, \qquad \Gamma_\zeta = 0.0025~\mbox{GeV}\,, \qquad \alpha = 1.10 \,. 
\eeqa
For those numbers, we have the following model prediction
\beqa
&&\Delta \overline{m}^{\rm mod}_s = 17.23~\mbox{ps}^{-1}\,,
\qquad
 \Delta \Gamma_s^{\rm mod}  = 0.125~\mbox{ps}^{-1}\,,
\qquad
 \overline{\Gamma}^{\rm mod}_s = 0.690~\mbox{ps}^{-1}\,,  \nonumber \\
&& 
\quad \tan{\phi_s^{\rm sl\,mod}} = -0.70\,,
\qquad \quad
\beta_s^{J/\psi \Phi\,{\rm mod}} = 0.13\,,
\eeqa
which has good agreement with the experimental measured values and has a total $\chi^2 = 3.2$ compared to $\chi^2 = 14.0$ in the SM. 
We also report the charge symmetry as $a^{s\,{\rm mod}}_{\rm sl} = - 5.0\times 10^{-3}$ for this point. 

\begin{figure}[!ht]
\begin{center}
\includegraphics[width=0.48\textwidth]{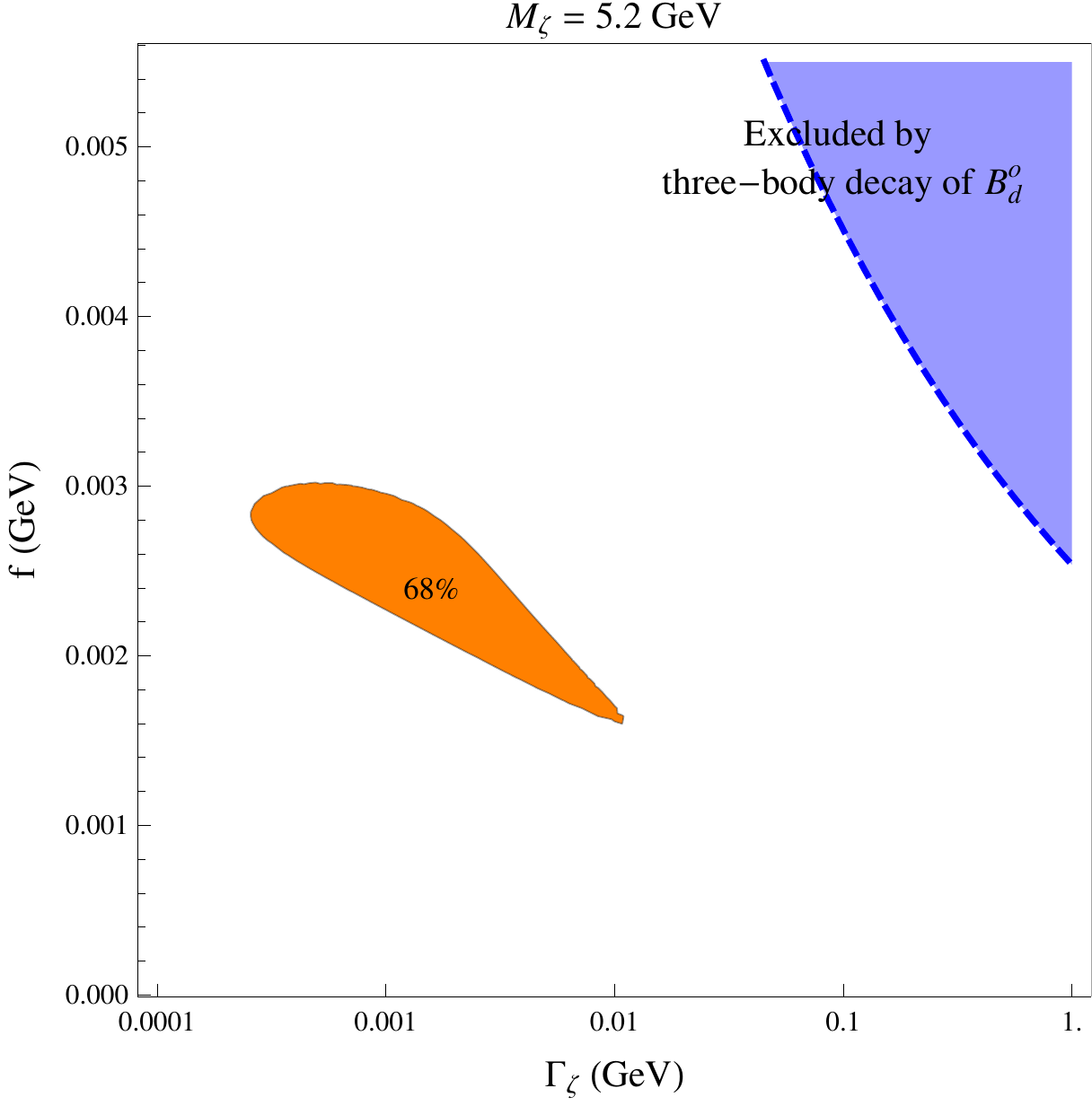} \hspace{2mm}
\includegraphics[width=0.48\textwidth]{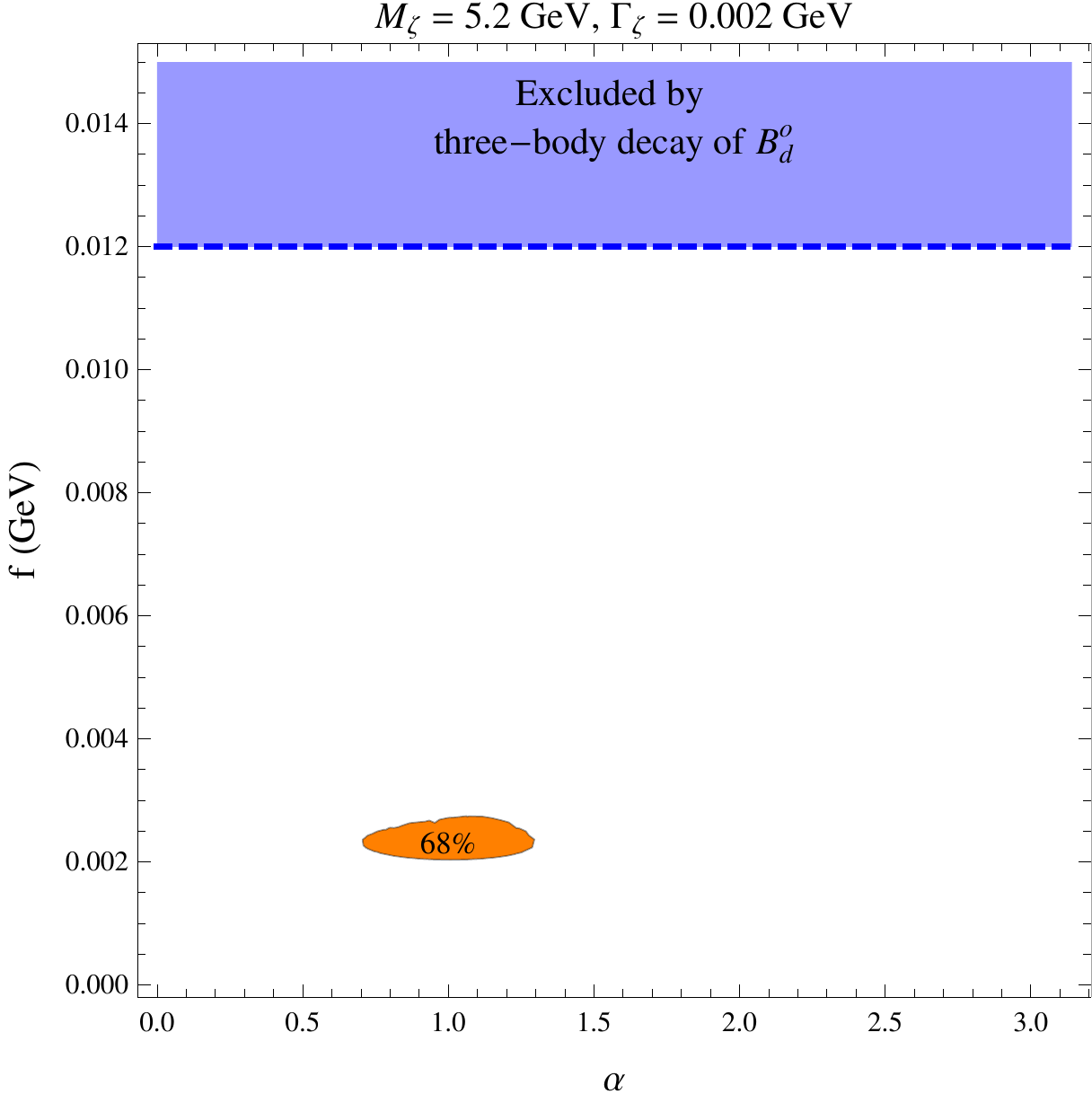}
\caption{Left panel: the best-fit region in the $\Gamma_\zeta$ and $f$ space for a fixed mass $M_\zeta=5.2$~GeV. The orange contour has $68\%$ C.L. after minimizing $\chi^2$ in terms of $\alpha$. The best fit has $\chi^2=2.0$. The blue region is excluded by requiring the three-body decay width  below the error of $\delta\Gamma_{B_d} =0.09~\mbox{ps}^{-1}$. Right panel:  the best-fit region in the $\alpha$ and $f$ space for $M_\zeta=5.2$~GeV and $\Gamma_\zeta=0.002$~GeV. The best fit has $\chi^2=2.3$.}
\label{fig:massiveMass}
\end{center}
\end{figure}

We present the best-fit region in the $\Gamma_\zeta$ and $f$ plane by fixing a specific $\zeta$ mass $M_\zeta = 5.2$~GeV in Fig.~\ref{fig:massiveMass}.
In the left panel of this figure, we still treat $\alpha$ as a floating parameter. The best-fit region prefers $\Gamma_\zeta$ within $10^{-4}-10^{-2}$~GeV. The best-fit region is not ruled out by the three-body decay width of $B_d$. In this plot and to have weak coupling between $\zeta$ and its decay  products, we don't extend the width of $\zeta$ to be above $1$~GeV, which is around 20\% of its mass. In the right panel of this figure, we present the allowed region in $\alpha$ and $f$ for fixed values of $M_\zeta=5.2$~GeV and $\Gamma_\zeta=0.002$~GeV.

\section{Discussion of New Meson Decay Modes and Conclusions}
\label{sec:conclusion}

From the best-fit region in the left panel of Fig.~\ref{fig:massiveMass}, the width of this new pseudoscalar particle should be above $10^{-4}$~GeV. We have only considered the total width constraint on various B meson decays so far. For some decay products from $\zeta$, more stringent bounds may be applied. Again, considering only two-body decays of $\zeta$ to SM particles and neglecting the final state masses, the coupling $\lambda_\zeta$ of $\zeta$ to SM particles should be above $\sim 0.1$.

We first  consider the situation where the $\zeta$ directly decays into two SM particles. Due to the strong constraints on branching ratios of $B_d$ to $K^0$ plus $e^+e^-$, $\mu^+\mu^-$, $\nu \bar{\nu}$, $\pi^+\pi^-$, $K^-\pi^+$ and $K^+K^-$, we are left with options $\zeta\rightarrow e^+\tau^-$, $\mu^+\tau^-$, $\tau^+\tau^-$, $\bar{c}u$, and $\bar{c}c$. The anomalous magnetic dipole moments $(g_\ell-2)/2$, which is approximately $\lambda_\zeta^2/(4\pi^2)(m_\tau^2/M^2_\zeta)^2$, are constrained to be smaller than $\sim 10^{-11}$, $\sim 10^{-9}$, and $\sim 10^{-2}$ for the electron, muon and tau, respectively. So, the two channels $\zeta\rightarrow e^+\tau^-$, $\mu^+\tau^-$ are ruled out. For the $\tau^+\tau^-$ channel, which is also pointed out recently in Ref.~\cite{Dighe:2010nj}, one may worry about  direct searches, for example at LEP II. While the production cross section $e^+e^-\rightarrow \tau^+ \tau^- \zeta\rightarrow 2 \tau^+ 2 \tau^-$ with $M_\zeta =5.0$~GeV and $\lambda_\zeta=0.1$ is $\sim 1$~fb and the total luminosity at LEP II is around 0.7~fb$^{-1}$. One may also worry about the modification on the width of the $Z$ boson decays. The decay width of $\Gamma(Z\rightarrow \tau^+\tau^- \zeta) \approx 0.013$~MeV for $M_\zeta =5.0$~GeV and $\lambda_\zeta=0.1$, which is also below the measurement error of $m_Z = 91.1876\pm0.0021$~GeV. Therefore, we conclude that the $\tau^+\tau^-$ channel is allowed.

The $\Delta C=1$ couplings $\zeta\,\bar{c}\,u$ can  contribute to an effective  $\Delta C=2$ operator which is highly constrained by $\overline{D}^0-{D}^0$ mixing, giving rise to a mass difference of order
\beqa\Delta m_D\sim \lambda_\zeta^2(f_D^2 m_D)/M^2_\zeta ,\eeqa  
compared with the measured value of~\cite{DDbarmixing}
\beqa
\Delta m_D=(1.6\pm0.5)\times10^{-14}~{\rm GeV}\ ,
\eeqa 
which constrains $\lambda_\zeta$ to be less than $\sim 2 \times 10^{-6}$.
We also note that if the coupling of $\zeta\,\bar{c}\,u$ is $CP$-violating,  the constraint  from   $D^0$ and $\overline{D}^0$ mixing is even more constraining. 

For the $\zeta\,\bar{c}\gamma_5 c$ couplings, we have mixing of $\zeta$ with the $\eta_c $ which gives rise to $\delta m_{\eta_c}
\approx 0.3$~MeV, which is below the experimental error $m_{\eta_c}=2980.5\pm 1.2$~MeV. After fragmentation, we anticipate $\zeta \rightarrow D\overline{D} (\pi^\prime s)$. From spectator decays, ${\rm Br}(B_s \rightarrow D_s^-D_s^+)$, ${\rm Br}(B_d \rightarrow D^-D_s^+)$ and ${\rm Br}(B_s \rightarrow J/\psi\,\eta)$ are estimated to be around $10^{-4}$ in our model, which is below but close to the current experimental errors~\cite{Esen:2010jq, Zupanc:2007pu, Adachi:2009usa}. So, we have two allowed two-body decay channels of $\zeta$: $\tau^+\tau^-$ and $\bar{c}c$. 

Also possible are three body decays, e.g. $\zeta\rightarrow c \bar u a$, with $a$ a new  stable particle. If $a$ differs from its own antiparticle then it is possible to avoid an excessive contribution to $\Delta C=2$ operators. We anticipate $\zeta \rightarrow D (\pi^\prime s) X$ with $X$ denoting missing particles. Making a three-body decay consistent with a large enough $\Gamma_\zeta$  is difficult however, due to phase space constraints.

A third possibility is to have $\zeta$ first decay into other lighter particles in the hidden sector and then have those lighter  particles decay back to the SM particles. If $\zeta$ decays into two identical particles $a$, more possible channels are open for $a$ decay. The $a$ field can decay into $e^+\tau^-$ and $\mu^+\tau^-$ if its couplings $\lambda_a \lesssim 10^{-4}$ to evade the constraints from the anomalous magnetic dipole moment. The $c\tau$ of the $a$ field is then estimated to be above $\sim 1~\mu m$. Although the modes to $\tau^+\tau^-$ and $\bar{c}c$ are kinematically forbidden because we need $M_a < M_\zeta/2$, the mode $\bar{c}u$ is allowed.  However for such a decay to take place without a significant displaced vertex requires a coupling which is too large to be consistent with $D^0-\overline{D}^0$ mixing, unless $a$ is not its own antiparticle.     The decays $a\rightarrow \pi^+\pi^-$ or $K^+K^-$ are also allowed because the exclusive searches for decay products of $B_d$ have not yet covered this kind of high-multiplicity final states like $B_d \rightarrow K^0\pi^+\pi^-\pi^+\pi^-$. Decays into $K^-\,\pi^+$  are allowed, but if $a$ is its own antiparticle then avoiding a  $\Delta S=2$ contribution to $K^0-\overline{K}^0$ mixing will imply a  significant  displaced vertex. 

We also have the option of $\zeta$ decay into two different light   particles $a_1$ and $a_2$. Now, we can have more combinations of final state particles with $a_1$ and $a_2$ decaying into different SM model particles. More specially, we can have one particle $a_2$ to be semi-stable and missing particle if its couplings to SM particles are weak enough to escape the detector. However, we can not allow both $a_1$ and $a_2$ to be (semi)stable particles because of the constraint from ${\rm Br}(B_d\rightarrow K^0\bar{\nu}\nu)<1.6\times 10^{-4}$. 

A fifth possibility, which may be difficult to constrain,   is that  $\zeta$ decays into    $ 2 a_1$    followed by decays of  the $a_1$ particle into SM particles and a hidden particle $a_2$, where $a_2$ may be semistable and escape the detector.  

We summarize the simplest allowed decay modes of $\zeta$ in Table~\ref{tab:decaymodes}.
\begin{table}[!ht]
\renewcommand{\arraystretch}{1.8}
\begin{center}
\begin{tabular}{c|c}
\hline \hline
   &  Decay  Modes     \\  \hline
Direct decay &  $\tau^+\tau^-$, $D \bar{D} ( \pi^\prime s)$, $D(\pi^\prime s) X$
   \\ \hline 
$\zeta\rightarrow 2\,a$   & $2\tau^+2e^-$, $2\tau^+2\mu^-$, $2D^+2\pi^-$, $2\pi^+2\pi^-$, $2\pi^+2\pi^-$, $2K^-2\pi^+$, $2K^+2K^-$
      \\ \hline
      $\zeta\rightarrow a_1+a_2$   &$X$ +  ($\tau^+e^-$, $\tau^+\mu^-$, $D^+\pi^-$, $\pi^+\pi^-$, $\pi^+\pi^-$, $K^-\pi^+$, $K^+K^-$)
      \\ \hline
   \hline
\end{tabular}
\end{center}
\caption{ Some allowed decay modes of $\zeta$, with $X$ representing missing particles. Other possible final states with different combinations of charges are also allowed. For  example in the $\zeta\rightarrow a_1+a_2$ case, different combinations of final states in the parenthesis are also allowed and not shown here. }
\label{tab:decaymodes}
\end{table}%
The new decay modes should account for approximately 3.5\% of the total width of the $B_s$ using the best-fit region in Fig~\ref{fig:massiveWidth}. So, if future experimental results find those new decay modes of $B_s$ in Table~\ref{tab:decaymodes} but not for $B_d$, our prediction for this new state $\zeta$ would be confirmed.

Before we end our paper, we make comments about the second operator in Eq.~(\ref{eq:operators}). This operator can induce flavor changing neutral current type decays of the top quark: $t \rightarrow c + \zeta$. Depending on the final states of $\zeta$ decay, there may exist a decay channel like $t \rightarrow c + \tau^+ \tau^-$, with the invariant mass of the $\tau$ pair around 5 GeV.  The branching ratio of this new decay channel is calculated as
\beqa
{\rm Br}(t \rightarrow c + \zeta) \,=\, \frac{|g^{tc\, 2}_A|\,m_t^3}{16\,\pi\,F^2\,\Gamma_t}\,\approx\,1.0\times 10^{-7}\times \left( \frac{10^6\,\mbox{GeV}}{F/|g^{tc}_A|}  \right)^{2}\,.
\eeqa
From Eq.~(\ref{eq:fandF}) and Fig.~\ref{fig:massiveWidth}, we see that if $g^{tc}_A\sim g^{bs}_A$ the explanation of anomalies in the $B_s$ symmetry in the familon framework predicts this branching ratio should be around $10^{-7}$, which may be tested at the LHC. 

In summary, we have shown that several anomalies in the $B_s$ system can be explained simultaneously if there is a new pseudoscalar, the $\zeta$,  mixing with $B_s$ mesons. From fitting to the five observables: $\Delta \overline{m}_s$, $\Delta\Gamma_s$, $\overline{\Gamma}_s$, $\phi_s^{\rm sl}$ and $\beta_s^{J/\psi \Phi}$, this pseudoscalar   is predicted to have a mass around $5$~GeV and a width around $10^{-3}$~GeV. Many viable decay modes can be found for the new pseudoscalar. As a result, the $B_d$, and $B^\pm$  should have non-negligible branching ratios of around $10^{-4}$ into a kaon  (or other fragmentation modes of $s \bar{d}$ and $s\bar{u}$) plus  the decay modes   of the  $\zeta$, such as those  listed in Table~\ref{tab:decaymodes}. Likewise the  $B_s$ should have a branching fraction of around $10^{-4}$ into $s\bar{s}$ fragmentation states such as $\phi$ or $\eta$, plus  the decay modes   of the  $\zeta$.  In addition, the  $B_s$ branching fraction into the $\zeta$ decay modes (with no additional particles) should be about $0.035$. Our model can be tested at the Tevatron and the LHC, particularly LHCb.  We also emphasize that $\phi_s^{\rm sl}\neq-2\beta_s^{J/\psi \Phi}$ in our model, in constrast with models in which new physics contributed only to $m^{12}_s$. More precise measurements of those two quantities would therefore distinguish our explanation of anomalies in the $B_s$ system from other approaches. Our motivation for this model is to better fit experimental data, and we have not attempted to justify the naturalness of the model or to discuss  the theoretical implications of such a new particle or particles, other than to make sure the model can be consistent with other experimental results.  However we note that light pseudoscalars  resulting from spontaneously broken approximate symmetries are common in models of flavor physics and/or dynamical electroweak symmetry breaking at the weak scale, as well as in models with new strongly coupled hidden sectors, which have a variety of theoretical motivations. It may seem somewhat surprising or coincidental that an exotic  hidden  meson should have a mass which is rather close to that of the $B_s$, however this is the simplest  viable way we are aware of  to get a large contribution to $\Gamma^{12}_s$.

\subsection*{Acknowledgements} 
We would like to thank Bogdan Dobrescu, Patrick Fox, Elvira Gamiz, Yuval Grossman,  Roni Harnik, Alex Kagan, Adam Martin, Michele Papucci and Andrew Wagner for useful discussions. We also thank the Aspen Center of Physics where part of this work was finished. This work (AN) was partially supported by the DOE under contract DE-FGO3-96-ER40956. Fermilab is operated by Fermi Research Alliance, LLC, under Contract DE-AC02-07CH11359 with the United States Department of Energy. 


\providecommand{\href}[2]{#2}\begingroup\raggedright\endgroup

\end{document}